\documentclass[smallextended]{svjour3}       % onecolumn (second format)
\smartqed  % flush right qed marks, e.g. at end of proof
\usepackage{graphicx}
\usepackage{amsfonts,amssymb,amscd,amsmath,enumerate,verbatim,calc}
\usepackage[colorlinks,citecolor=blue]{hyperref}
\usepackage{color}

\spnewtheorem{result}{Result}{\bf}{\it}
\begin{document}

\title{On Einstein equations with cosmological constant in braneworld models }
%placed here. General acknowledgments should be placed at the end of the %article.}
%\subtitle{On warped product spaces with cosmological constant}

\titlerunning{ Reduction of Einstein equations with cosmological constant}        % if too long for running head

\author{F. Gholami,  F. Darabi,  A. Haji-Badali}

\authorrunning{F. Gholami \and F. Darabi    \and  M. Faghfouri   \and
        A. Haji Badali } % if too long for running head
\institute{F. Gholami (Corresponding author)\at
           Department of Mathematics, Faculty of Basic Sciences,   University of Bonab, Bonab, Iran\\
              \email{fateme.gholami@bonabu.ac.ir}
               \and
                F. Darabi  \at
           Department of Physics, Azarbaijan Shahid Madani University, Tabriz, Iran\\
              \email{f.darabi@azaruniv.edu}
                \and
        A. Haji-Badali  \at
              Department of Mathematics, Faculty of Basic Sciences,University of Bonab, Bonab, Iran\\
              \email{haji.badali@bonabu.ac.ir}
}

\date{Received: date / Accepted: date}

\maketitle

\begin{abstract}
In this paper, we investigate the Einstein equations with
cosmological constant for Randall-Sundrum (RS) and
Dvali-Gabadadze-Porrati (DGP) models to determine the warp
functions in the context of warp product spacetimes. In RS model, it is shown that Einstein's equation in the bulk
is reduced into the brane as a vacuum equation, having vacuum solution, which is not affected by the cosmological constant in the bulk. In DGP model, it is shown that the Einstein's equation in the bulk is reduced into the brane and along the extra dimension, where both equations are affected by the cosmological constant in the bulk. We have solved these equations in DGP model, subject to vanishing cosmological constants on the brane and along extra dimension, and obtained exact solutions for the warp functions. The solutions depend on the typical values of cosmological constant in the bulk as well as the dimension of the brane. So, corresponding to the typical values, some solutions have exponential behaviours which may be set to represent {\it warp inflation} on the brane,
and some other solutions have oscillating behaviours which may be set
to represent {\it warp waves} or {\it branes waves} along the extra dimension.
\end{abstract}
\maketitle

\section{Introduction}

In recent years, the modification of Einstein equations has received a
large amount of attention \cite{f(R)}, specially in models with
spacetime dimensions other than four, like RS models \cite{RS}, \cite{Randall.SundrumII} and DGP model \cite{DGP}. Many of these
spacetimes may be considered as base conformal warped product
spacetimes for which metrics are in the form of a mixture of a
conformal metric on the base and a warped metric. These metrics
are very familiar in different theories of physics, such as
general relativity theory, extra-dimensional theories, string theory, super gravity and quantum
gravity. Finding the viable and correct form of Einstein equations
is one of the most important challenges in any general
relativistic theory of gravitation including the modified models
of gravity and higher dimensional theories of gravity.

One of the important motivations of generalized Einstein gravity
models is the construction of an appropriate quantum gravity model. It
is shown that the AdS/CFT correspondence is an extremely useful
tool for investigation of gravity from geometric point of view
which helps one to find different aspect of quantum gravity. Moreover, it turns out that there is a kind of complementarity between
AdS/CFT and RS scenarios and that they are locally similar,
while their global geometries are different \cite{McInnes} (see
also \cite{Gubser,Verlinde,Chan,Hawking,DuffPRL,DuffJMP} for more
details). Such a relation motivates one to consider and study the RS models of gravitation. The RS models are based on a braneworld theory developed to solve the hierarchy problem of the Standard Model, namely weakness of gravity relative to the other fundamental forces of nature. This theory includes a finite five-dimensional warped bulk with two branes which are separated by a large fifth dimension. One brane has positive energy, and the other
brane has negative energy. The RS1 model, first of the two models, has a finite size for the extra dimension between the two branes, whereas in the
second RS2 model, the second brane has been placed infinitely far away from the first brane, so that there is actually only one brane left in the model.

On the other hand, DGP gravity (known as ``brane-induced gravity), introduces a new and interesting way of modification to gravity at large distances, capable of producing an accelerated expansion of the universe without a nonvanishing
cosmological constant \cite{Deffayet1}-\cite{Deffayet2}. There are different motivations of considering the DGP braneworld
scenario in the context of gravitation and cosmology. It is shown
that the soft massive gravity predicted in the DGP model or its
extensions is free of Boulware-Deser ghost \cite{DGP2,DGP3}. In
addition, one of the most important implications of the DGP is the
existence of self-accelerating solutions (for more details related
to the cosmology and phenomenology of DGP, see
\cite{Deffayet1,Deffayet2,Lue1,Lue2,Lue3}). Based on the
phenomenology of DGP, the universe is a $4-$dimensional spacetime
brane embedded in a five-dimensional flat bulk. All particles and
fields are constrained to remain on the brane while gravity is
free to explore the empty bulk. One may also recover the usual gravitational interaction at small distances using the scalar
sector of nonlinearities of the DGP model which leads to observable signature and precision tests of gravity at solar system scale.

The main idea of braneworld gravity is that the visible universe is restricted to a three-dimensional brane inside a higher-dimensional bulk space. In the braneworld gravity, some of the extra dimensions are extensive and the branes may move through this bulk, interact with the bulk, and possibly with other branes. These interactions may affect our brane and hence introduce effects which are not predicted by standard cosmological models. Extensions of the braneworld gravity with supersymmetry in the bulk may be also promising in addressing the well known cosmological constant problem.

The $5$-dimensional brane world theories like RS \cite%
{RS}, \cite{Randall.SundrumII} and DGP models \cite{DGP}, may
be considered as warped product spacetimes. We have already
obtained, in the context of warped product spacetimes, the
Einstein equations with cosmological constant in arbitrary $(m+n)D$ and $(1+n)D$ multidimensional spacetime with multiply-warped product metric $(\bar{M},\bar{g})$
 \cite{FGH-FD-AH}, \cite{MF-AH-FGH}.  In this paper, motivated by the previous works, we consider arbitrary multidimensional brane world metrics of RS and DGP models with a cosmological constant, as warped product spacetimes. Note that, unlike the original RS and DGP models, here
we suppose that the only nonvanishing energy-momentum tensor is that of a higher dimensional cosmological constant $\bar{\Lambda}$. Then, we show that the Einstein equations $\bar{G}_{AB}=-\bar{\Lambda}\bar{g}_{AB}$ with a cosmological
constant $\bar{\Lambda}$ are reducible to the Einstein equations
on the brane and in the bulk with reduced cosmological constants
which are determined in terms of the warped functions. In this way, both
Einstein tensor and energy-momentum tensor are given in terms of
the warped functions and their space-time dynamics.

The paper is organized as follows. In the next section, we describe the warp
product structure of RS metric and obtain the reduced Einstein equations on the brane and in the bulk with reduced cosmological constants. In section 3, we describe the warp product structure of DGP metric and obtain the reduced
Einstein equations on the brane and in the bulk with reduced cosmological
constants. We end the paper with a brief conclusion.

\section{RS model}

In this section, we intend to consider the multidimensional RS braneworld cosmological model in the absence of matter and brane curvature terms. We only consider a multidimensional cosmological constant in the RHS of multidimensional Einstein equations, in the bulk, and obtain the reduced Einstein equations on the brane and along the extra dimensions. In this regard, we shall define the geometry of RS model as a warped product metric. 

Let $\bar{M}=I\times
_{f}M$ be a warped product manifold, in which $I$ is an open
interval of the real line $\mathbb{R}$, $M$ is a Riemannian
$n$-manifold endowed with the Lorentzian metric and $f=e^{-2k|y|}$
is a positive smooth function on $M$. Also, suppose that $\bar{g}$
is a pseudo-Riemannian metric on $\bar{M}$ defined by
\begin{equation}
\bar{g}={e^{-2|y|/\ell }}g_{\alpha \beta }{dx^{\alpha }}dx^{\beta }+dy^{2},
\label{eq: metric1}
\end{equation}%
with the following related line element%
\begin{equation}
ds^{2}={e^{-2|y|/\ell }}g_{\alpha \beta }dx^{\alpha }dx^{\beta }+dy^{2},
\end{equation}%
where $0\leq y\leq \pi r_{c}$ is the extra-dimensional coordinate
and $r_{c}$ is essentially a compactification \textquotedblleft
radius and $\ell $ is a non-zero constant denoting the curvature
radius of $AdS_{5}$. The mentioned metric $\bar{g}$ is called as
the RS metric given by its local
components%
\begin{eqnarray}
\bar{g}_{\alpha \beta } &=&e^{-2|y|/\ell }g_{\alpha \beta }, \\
\bar{g}_{yy} &=&1,  \label{eq:metric2} \\
\bar{g}_{\alpha y} &=&0,
\end{eqnarray}%
where $g_{\alpha \beta }(x^{\mu })$ are the local components of $n-$%
dimensional manifold $g$ (i.e.: $\alpha ,\beta ,...\in
\{1,...,n\}$). Now, we recall some basic facts from the Ref.
\cite{ONeill}.

Considering $\bar{M}=I\times _{f}M$ as a Lorentzian warped product of $(M,g)$, we can obtain \cite{ONeill}%
\begin{align}
\bar{R}_{yy}& =-4n\ell ^{-2},  \label{eq:Ricij} \\
\bar{R}_{y\alpha }& =0,  \label{Rya} \\
\bar{R}_{\alpha \beta }& ={R}_{\alpha \beta }-(4n\ell ^{-2})\bar{g}_{\alpha
\beta },  \label{Rab}
\end{align}%
where the components of $R_{\alpha \beta }=Ric^{M}(\partial _{\alpha
},\partial _{\beta })$ are the local components of Ricci tensor of $(M,g)$.

In addition, regarding $\bar{M}=I\times _{f}M$ as a warped product manifold
and $f=e^{-2|y|/\ell }$, the scalar curvature $\bar{S}$ of $(\bar{M},\bar{g}%
)$ admits the following relation \cite{ONeill}%
\begin{equation}
\bar{S}=-8n\ell ^{-2}+\frac{S^{M}}{e^{-2|y|/\ell }}-4n(n-1)\ell ^{-2},
\label{Sbar}
\end{equation}%
in which $S^{M}$ is the scalar curvature of $(M,g)$.

Moreover, assuming $\bar{G}=\bar{Ric}-\frac{1}{2}\bar{S}\bar{g}$ as the
Einstein gravitational tensor field of $(\bar{M},\bar{g})$, and according to
Eqs. (\ref{eq:Ricij})-(\ref{Rab}) and (\ref{Sbar}), one can find the
following equation%
\begin{align}
\bar{G}_{yy}& =2n(n-1)\ell ^{-2}-\frac{S^{M}}{2e^{-2|y|/\ell }},
\label{eq:ein1} \\
\bar{G}_{y\alpha }& =0,  \label{eq:ein2} \\
\bar{G}_{\alpha \beta }& =G_{\alpha \beta }-\left( 2n(n-1)\right) \bar{g}%
_{\alpha \beta },  \label{eq:ein3}
\end{align}%
where the components of $G_{\alpha \beta }$ is the local components of the
Einstein gravitational tensor field of $(M,g)$.

Now, we consider Einstein's equations with the cosmological constant $\bar{%
\Lambda}$ on $(\bar{M},\bar{g})$ as $\bar{G}_{AB}=-\bar{\Lambda}\bar{g}_{AB}$
which are equivalent to the following equations%
\begin{align}
& \bar{\Lambda}=-2n(n-1)\ell ^{-2},  \label{Lambdabar} \\
& G_{\alpha \beta }=0,  \label{Gab} 
\end{align}
These results are straightforward consequences of Eqs. (\ref{eq:ein1})-(\ref%
{eq:ein3}) with Eq. (\ref{eq:metric2}). Moreover,
we note that $G_{yy}=0$ because of $g_{yy}=1$. 

It turns out that a nonvanishing cosmological constant $%
\bar{\Lambda}$ is necessarily negative for $n>1$ and, for a given dimension $%
n$, its order of magnitude is determined by the curvature radius $\ell $ of
the brane. It is clear that as the curvature radius $\ell $, namely the
characteristic length in the exponent of the warp factor, becomes smaller (a
fast decaying warp function), the order of magnitude of the cosmological
constant $\bar{\Lambda}$ becomes larger. Another result is that the warp
factor is related, through the curvature radius $\ell $, to the cosmological
constant $\bar{\Lambda}$ and the dimension $n$.

Moreover, it turns out that the induced Einstein equations on the
brane is the vacuum equation. This means that the cosmological
constant $\bar{\Lambda} $ in the bulk induces a vanishing
cosmological constant on the brane. In other words, the Einstein
equations on the brane is free of the cosmological constant in the
bulk and so it is not affected by the curvature radius $\ell $ of
$AdS_{5}$.

\section{DGP cosmological model}

In this section, we intend to consider the special case of multidimensional metric of DGP braneworld cosmological model, studied by Deffayet \cite{Deffayet1}, in the absence of matter and brane curvature terms. We write the multidimensional
Einstein equations and obtain the reduced Einstein equations on the brane and along the extra dimensions. 

DGP braneworld cosmological model is described by the following action
\begin{equation}
S_{(5)}=-\frac{1}{2}\int d^5 X \sqrt{-\bar{g}}\bar{R}+\int d^5 X \sqrt{-\bar{g}}\mathcal{L}_m-\frac{1}{2}\int
d^4 X \sqrt{-{g}}{R}.
\end{equation}
The first integral corresponds to the Einstein-Hilbert action in five dimensions
for a five-dimensional metric $\bar{g}_{AB}$ in the bulk metric, $\bar{R}$ being its scalar curvature. The second integral corresponds to the matter action in five dimensions. The third integral corresponds to the Einstein-Hilbert action for the induced metric $g_{ab}$ on the brane, $R$ being its scalar curvature. Variation of the action with respect to the metric $\bar{g}_{AB}$
gives the Einstein equations
\begin{equation}
\bar{G}_{AB}=\bar{S}_{AB},
\end{equation}
where $\bar{G}_{AB}$ is the five-dimensional Einstein tensor, and the tensor $\bar{S}_{AB}$ is the sum of the energy momentum tensor of matter and the contribution of the brane scalar curvature. 

Unlike this model, here we consider a multidimensional cosmological model with the Einstein-Hilbert action having a cosmological constant $\bar{\Lambda}$ described by
\begin{equation}
S_{(5)}=-\frac{1}{2}\int d^5 X \sqrt{-\bar{g}}(\bar{R}+2\bar{\Lambda}).
\end{equation}
Variation of this action gives the multidimensional Einstein equations
\begin{equation}
\bar{G}_{AB}=-\bar{\Lambda}\bar{g}_{AB}.
\end{equation}
In the following, we shall define the geometry of DGP braneworld cosmological model as a multidimensional
warped product metric. 

Let $\bar{M}=I\times _{f_{1}}M_{1}\times_{f_{2}}K$ where both $K$ and $I$ are $1$-dimensional manifolds and $(M,g)$ is a Riemannian $m$-manifold endowed with the Lorentzian metric,
 and $f_{1}$ and $f_{2}$ are, respectively, $%
a(\tau ,y)$ and $b(\tau ,y)$ to obtain $\bar{g}$ as a pseudo-Riemannian metric on $\bar{M}$ defined by
\begin{equation*}
\bar{g}=-d\tau ^{2}+a^{2}(\tau ,y)g_{\alpha \beta }dx^{\alpha }dx^{\beta
}+b^{2}(\tau ,y)dy^{2},
\end{equation*}%
where $\tau $ is the
comoving time coordinate and $y$ is the coordinate of the $m^{th}-dimension$
which is compact (i.e.: its interval $I$ is chosen to be $-1/2\leq y\leq 1/2$ with two endpoints).
This metric gives us the following $m$-dimensional line element\footnote{This
line element is the special case of DGP braneworld cosmological model in
which the laps function is set to $n(\tau ,y)=1$.}
\begin{equation}
ds^{2}=-d\tau ^{2}+a^{2}(\tau ,y)g_{\alpha \beta }dx^{\alpha
}dx^{\beta }+b^{2}(\tau ,y)dy^{2},  \label{MetricDGP}
\end{equation}%
where its local components are given by 
\begin{align}
& \bar{g}_{\tau \tau }=-1,  \label{eq:metric1} \\
& \bar{g}_{\alpha \beta }=a^{2}(\tau ,y)g_{\alpha \beta }, \\
& \bar{g}_{yy}=b^{2}(\tau ,y),  \label{eq:metric3} \\
& \bar{g}_{y\alpha }=0,\\ 
& \bar{g}_{\tau \alpha }=0.
\end{align}%
Like previous sections, we get Ricci tensor and Einstein gravitational field
tensor on $\bar{M}$.

By considering $\bar{M}=I\times_{f_{1}}M_{1}\times _{f_{2}}K$ as a warped product manifold, we
can obtain \cite{Dobarro}
\begin{align}
\bar{Ric}_{\tau \tau }& =-\left( m\frac{a^{\prime \prime }+\ddot{a}}{a}+%
\frac{b^{\prime \prime }+\ddot{b}}{b}\right) ,  \label{eq:ric1} 
\end{align}
\begin{equation}
\bar{Ric}_{\alpha \beta } =Ric_{\alpha \beta }-\bar{g}_{\alpha \beta }\left[
\frac{a^{\prime \prime }+\ddot{a}}{a}+\left( \frac{a^{\prime 2}+\ddot{a}^{2}%
}{a^{2}}\right) \times (m-1)+m\frac{(a^{\prime }+\dot{a})(b^{\prime }+\dot{b})}{ab}\right],  \label{eq:Ric2} 
\end{equation}
\begin{align}
\bar{Ric}_{yy}& =-\left( \frac{b^{\prime \prime }+\ddot{b}}{b}+\frac{%
(a^{\prime }+\dot{a})(b^{\prime }+\dot{b})}{ab}\right) b^{2},
\label{eq:ric3} \\
\bar{Ric}_{\tau \alpha }& =0, \\
\bar{Ric}_{\alpha y}& =0.
\end{align}%
where a prime $'$ and a dot $^.$ denote the derivatives with respect to $y$
and $\tau$, respectively, and the components $Ric_{\alpha \beta }$ are the local components of Ricci
tensor of $(M,g)$.

Moreover, regarding $\bar{M}=I\times _{f_{1}}M_{1}\times _{f_{2}}K$ as a
multiply-warped products manifold, one finds that the scalar curvature $S$
of $(\bar{M},\bar{g})$ admits the following expression\textbf{\ }\cite%
{Dobarro}
\begin{equation}
\bar{S} =-2\left( m\frac{a^{\prime \prime }+\ddot{a}}{a}+\frac{b^{\prime
\prime }+\ddot{b}}{b}\right) +\frac{S^{M}}{a^{2}}-m(m-1)\frac{a^{\prime 2}+\dot{a}^{2}}{a^{2}}-m\frac{(a^{\prime }+\dot{a})(b^{\prime }+\dot{b})}{ab},  \label{eq:Scalar}
\end{equation}%
where $S^{M}$ is the scalar curvature of $(M,g)$.

In addition, considering $\bar{G}$ as the Einstein gravitational tensor
field of $(\bar{M},\bar{g})$ with the corresponding Einstein gravitational
field tensor $\bar{G}=\bar{Ric}-\frac{1}{2}\bar{S}\bar{g}$ and using Eqs. (%
\ref{eq:metric2}) and (\ref{eq:Scalar}), we have the following equations%
\begin{equation}
\bar{G}_{\alpha \beta }=G_{\alpha \beta }+\bar{g}_{\alpha \beta }\left[
(m-1)(\frac{m}{2}-1)\frac{a^{\prime 2}+\dot{a}^{2}}{a^{2}}+ m(\frac{a^{\prime \prime }+\ddot{a}}{a}+\frac{(a^{\prime }+\dot{a}%
)(b^{\prime }+\dot{b})}{ab})\right], 
\end{equation}
\begin{equation}
\bar{G}_{yy}=b\left[ \frac{a^{\prime \prime }+\ddot{a}}{a}+\frac{m(m-1)}{2}%
(\frac{a^{\prime 2}+\dot{a}^{2}}{a^{2}})+(\frac{m}{2}-1)(\frac{(a^{\prime }+\dot{a})(b^{\prime }+\dot{b})}{ab%
})-\frac{S^{M}}{a^{2}}\right] ,
\end{equation}%
\begin{equation}
\bar{G}_{\tau \tau }=\frac{-1}{2}\left[ \frac{S^{M}}{a^{2}}+m(m-1)\frac{%
a^{\prime 2}+\dot{a}^{2}}{a^{2}}-m\frac{(a^{\prime }+\dot{a})(b^{\prime }+\dot{b})}{ab}\right], \end{equation}
\begin{align}
& \bar{G}_{\alpha \tau }=0, \\
& \bar{G}_{y\tau }=0, \\
& \bar{G}_{y\alpha }=0, 
\end{align}
where the components $G_{\alpha \beta }$ are the local components of the
Einstein gravitational field tensor $G$ of $(M,g)$.

As the final interesting result, we may consider the Einstein equations $%
\bar{G}_{AB}=-\bar{\Lambda}\bar{g}_{AB}$ on $(\bar{M},\bar{g)}$ with
cosmological constant $\bar{\Lambda}$ and also using the same strategy explained
in \cite{FGH-FD-AH}, we can obtain the following equations
\begin{equation}
\bar{\Lambda} =\frac{-1}{2}\left[ m(m-1)(\frac{a^{\prime \prime }+\ddot{a}}{%
a})+(m^{2}+1-m)\times \frac{(a^{\prime }+\dot{a})(b^{\prime }+\dot{b})}{ab}+m\frac{%
b^{\prime \prime }+\ddot{b}}{b}\right] ,  \label{barLambda} 
\end{equation}
\begin{equation}
\Lambda = a^{2}(1-\frac{m}{2})\left[ (m-1)\frac{aa^{\prime \prime }-{%
a^{\prime }}^{2}+a\ddot{a}-{\dot{a}}^{2}}{a^{2}}-\frac{m(a^{\prime }+\dot{a})(b^{\prime }+\dot{b})}{ab}+\frac{b^{\prime \prime }+\ddot{b}}{b}\right] ,  \label{Lambda} \end{equation}
\begin{equation}
\Lambda_{y} =\frac{b^{2}}{2}\left( m(\frac{a^{\prime \prime }+\ddot{a}}{a}%
)-(\frac{(a^{\prime }+\dot{a})(b^{\prime }+\dot{b})}{ab})\right),
\label{Lambday}
\end{equation}
\begin{align}
G_{\alpha \beta }& =-\Lambda g_{\alpha \beta }, \\
G_{yy}& =-\Lambda _{y}g_{yy}. \label{Gyy}
\end{align}

Keeping in mind the Einstein equations, $G_{\alpha \beta }=-\Lambda g_{\alpha \beta }$, and $G_{yy}=-\Lambda
_{y}g_{yy}$, and that $\bar{\Lambda}$, $\Lambda $, and $\Lambda _{y}$ are
cosmological constants, we can combine the above coupled partial
differential equations, (\ref{barLambda}), (\ref{Lambda}) and (\ref{Lambday}%
), and solve them to obtain the warp functions $a(\tau ,y)$ and $b(\tau ,y)$
by applying the appropriate boundary conditions. 

In this regard, we first note that $G_{yy}=0$ because of $g_{yy}=1$. So, considering Eq.(\ref{Gyy}), we have to impose $\Lambda _{y}=0$ which leads to the following equation
\begin{equation}\label{Gyy=0}
 m\frac{a^{\prime \prime }+\ddot{a}}{a}%
=\frac{(a^{\prime }+\dot{a})(b^{\prime }+\dot{b})}{ab}.
\end{equation}
Then, using (\ref{Gyy=0}) in (\ref{barLambda}) and (\ref{Lambda}), we obtain
the following equations
\begin{equation}
\bar{\Lambda} =\frac{-m}{2}\left[ (m^2+m)\frac{a^{\prime \prime }+\ddot{a}}{%
a}+\frac{b^{\prime \prime}+\ddot{b}}{b}\right],
\end{equation}
\begin{equation}
\Lambda = a^{2}(1-\frac{m}{2})\left[ (m-1)\frac{aa^{\prime \prime }-{%
a^{\prime }}^{2}+a\ddot{a}-{\dot{a}}^{2}}{a^{2}}- 
\left(m^2 \frac{a^{\prime \prime }+\ddot{a}}{%
a}-\frac{b^{\prime \prime }+\ddot{b}}{b}\right)\right].
\end{equation}
Now, considering $\bar{\Lambda}$ and $\Lambda$ as constants, these second order partial differential equations should be solved to obtain the warp functions $a(\tau ,y)$ and $b(\tau ,y)$. For the sake of simplicity, we assume
$\Lambda=0$ and rewrite these equations as follows
\begin{equation}\label{41}
\frac{2\bar{\Lambda}}{m}+(m^2+m)\frac{a^{\prime \prime }+\ddot{a}}{%
a} = -\frac{b^{\prime \prime}+\ddot{b}}{b},
\end{equation}
\begin{equation}
(m-1)\frac{aa^{\prime \prime }-{%
a^{\prime }}^{2}+a\ddot{a}-{\dot{a}}^{2}}{a^{2}} -m^2 \frac{a^{\prime \prime }+\ddot{a}}{a} = -\frac{b^{\prime \prime }+\ddot{b}}{b},
\end{equation}
which result in the following equation
\begin{equation}
(2m^2+1) \frac{a^{\prime \prime }}{a} + (m-1)\frac{a'^{2}}{a^2}+\frac{2\bar{\Lambda}}{m}=-\left[(2m^2+1)\frac{\ddot{a}}{a}
+(m-1)\frac{\dot{a}^2}{a^2}\right].
\end{equation}
By assuming $a(\tau ,y)=T(\tau)Y(y)$ and putting in the above equation we
obtain
\begin{equation}
(2m^2+1) \frac{Y^{\prime \prime }}{Y} + (m-1)\frac{Y'^{2}}{Y^2}+\frac{2\bar{\Lambda}}{m}=-\left[(2m^2+1)\frac{\ddot{T}}{T}
+(m-1)\frac{\dot{T}^2}{T^2}\right].
\end{equation}
Now, this equation can be divided into two following equations
\begin{equation}
(2m^2+1) \frac{Y^{\prime \prime }}{Y} + (m-1)\frac{Y'^{2}}{Y^2}=C,
\end{equation}
\begin{equation}
(2m^2+1)\frac{\ddot{T}}{T}
+(m-1)\frac{\dot{T}^2}{T^2}=-(C+\frac{2\bar{\Lambda}}{m}),
\end{equation}
where $C$ is an arbitrary a constant.
The solutions are obtained as
\begin{align}
Y \left( y \right) =& {2}^{-{\frac {2m^2+1}{m \left( 1+2\,m \right) }}}{{\rm e}^{-{\frac {\sqrt {Cm}y}{m\sqrt {1+2\,m}}}}}\times \nonumber \\
&\left( C {m}^{-
1} \left( {{\rm e}^{2\,{\frac {\sqrt {Cm}\sqrt {1+2\,m}\,y}{2\,{m}^
{2}+1}}}}{C_1}-{C_2} \right) ^{-2} \left( 1+2\,m \right) ^{-
1} \right) ^{-{\frac {2\,{m}^{2}+1}{2m \left( 1+2\,m \right) }}},
\end{align}
\begin{align}
T \left( \tau \right) =& {2}^{-{\frac {2m^2+1}{m \left( 1+2\,m \right) }}}{{\rm e}^{-{\frac {\sqrt {-(Cm+2\bar{\Lambda})}\,\tau}{m\sqrt {1+2\,m}}}}}\times \nonumber \\
&\left( - (C+2\bar{\Lambda}/m){m}^{-1}   \left( {{\rm e}^{2\,{\frac {\sqrt {-(Cm+2\bar{\Lambda})}\sqrt {1+2\,m}\,\tau}{2\,{m}^
{2}+1}}}}{C_1}-{C_2} \right) ^{-2} \left( 1+2\,m \right) ^{-
1} \right) ^{-{\frac {2\,{m}^{2}+1}{2m \left( 1+2\,m \right) }}},
\end{align}
where $C_1$ and $C_2$ are constants of integration. For simplicity, by setting
$C_1=1$ and $C_2=0$ we obtain
\begin{align}\label{a(t,y)}
a(\tau, y)=&A\,{{\rm e}^{-{\frac {(\sqrt {Cm}\,y+\sqrt {-(Cm+2\bar{\Lambda})}\, \tau)}{m\sqrt {1+2\,m}}}}}
\left(   {{\rm e}^{\,{-\frac {(\sqrt {Cm}\,y+\sqrt {-(Cm+2\bar{\Lambda})}\, \tau)\sqrt {1+2\,m}}{2\,{m}^
{2}+1}}}}  \right) ^{-{\frac {2\,{m}^{2}+1}{2m \left( 1+2\,m \right) }}},
\end{align}
where 
$$
A={2}^{-{\frac {2(2m^2+1)}{m \left( 1+2\,m \right) }}} \left(  -C (C+2\bar{\Lambda}/m) \right) ^{-{\frac {2\,{m}^{2}+1}{2m \left( 1+2\,m \right) }}}\times \left(   {m} \left( 1+2\,m \right) \right) ^{{\frac{m \left( 1+2\,m \right) }{2\,{m}^{2}+1}}}.
$$
By substituting (\ref{a(t,y)}) in Eq.(\ref{41}), we obtain 
\begin{equation}\label{50}
\frac{b^{\prime \prime}+\ddot{b}}{b} = C',
\end{equation}
where $C'=-{\frac {\bar{\Lambda}\, \left( 3+7\,m \right) }{2m \left( 1+2\,m
 \right) }}
$ is a constant. By assuming $b(\tau ,y)=\Theta(\tau)\Omega(y)$, we may divide Eq.(\ref{50})
into two equations
\begin{equation}\label{51}
\frac{\Omega^{\prime \prime}(y)}{\Omega(y)} = K,
\end{equation}
\begin{equation}\label{52}
\frac{\ddot{\Theta}(\tau)}{\Theta(\tau)} =C'-K,
\end{equation}
where $K$ is an arbitrary constant. By solving the above equations for
$\Omega$ and $\Theta$, we can easily find the following solution
\begin{equation}\label{b(t,y)}
b(\tau, y)=({\it C^{\prime}_1}\,{{\rm e}^{\sqrt {K}y}}+{\it C^{\prime}_2}\,{
{\rm e}^{-\sqrt {K}y}})
({\it C^{\prime\prime}_1}\,{{\rm e}^{\sqrt {C'-K}\tau}}+{\it C^{\prime\prime}_2}\,{
{\rm e}^{-\sqrt {C'-K}\tau}}),
\end{equation}
where $C^{\prime}_1$, $C^{\prime}_2$, $C^{\prime\prime}_1$ and $C^{\prime\prime}_2$
are constants of integration.

The solutions for warp functions ${a(\tau,y)}$ and ${b(\tau,y)}$ are capable of producing the exponential and/or oscillating solutions, according to the chosen values for $C, m, \bar{\Lambda}$ and $K$. The exponential solutions
may be set to describe inflationary behaviors of the warp functions, on the
brane, the inflation rates of which are determined by the chosen values for $C, m, \bar{\Lambda}$ and $K$; whereas the oscillating solutions may be set to describe wave-like behaviors, interpreted as {\it warp waves} or {\it brane waves} moving along the extra dimension $y$, the frequencies and velocities of which are determined by the chosen values for $C, m, \bar{\Lambda}$ and $K$.

\section{Conclusion}

In Space-Time-Matter (STM) theory, the reduction of Einstein equations in warped product metrics was studied and the following results were obtained. \cite{MF-AH-FGH}
\begin{itemize}
\item For $\bar{M}=M_{1}\times _{f}M_{2}$,
\begin{align}
& \bar{\Lambda}=\left( \frac{m+n-2}{2m}\right) \left( n\frac{\Delta f}{f}%
-S^{M_{1}}\right)  \\
& G_{\alpha \beta }=-f^{2}(1-\frac{n}{2})\left[ \frac{\Delta f}{f}(1-\frac{n%
}{m})+\frac{S^{M_{1}}}{m}+\right.  \nonumber \\
& \left. (n-1)\frac{{\left\Vert {\ grad\;f}\right\Vert }^{2}}{f^{2}}\right]
h_{\alpha \beta }~,
\end{align}

\item For $\bar{M}=I\times _{f}M$:
\begin{align}
& \bar{\Lambda}=\frac{1}{8}n(n-1)(B^{2}+2B^{\prime }),  \label{126} \\
& G_{\alpha \beta }=\left[ \frac{1}{4}(n-1)(n-2)f^{2}B^{\prime }\right]
h_{\alpha \beta }~,  \label{127}
\end{align}
\end{itemize}
where in both cases, it is seen that the reduced cosmological constants $%
\Lambda$ and the higher dimensional cosmological constant
$\bar{\Lambda}$ are dependent on each other.

In the present paper, we have studied RS and DGP multidimensional
warped product spacetimes where the Einstein equations are
supported merely by the cosmological constants in the bulk. We have
shown that these Einstein equations are reducible to the Einstein
equations with reduced cosmological constants on the brane and
along the extra dimension, which depend generally on the dimension
of the brane, warp functions and the scale factors on the brane
and the bulk. 

In RS model, we have shown that the Einstein
equations on the brane is a vacuum equation which is not affected
by the cosmological constant in the bulk. 
In DGP model, we have shown that the Einstein equations
with a cosmological constant in the bulk is reduced to the
Einstein equations with a cosmological constant on the brane and
the Einstein equations with a cosmological constant along the
extra dimension, where both of these reduced cosmological
constants are affected, in principle, by the cosmological constant in the bulk. We have argued that the cosmological constant along the
extra dimension have to be zero. For the sake of simplicity, we have also
assumed the cosmological constant on the brane to be zero, which affects
the cosmological constant in the bulk. Then, we have solved
the Einstein equations by assuming the separation of variables $\tau$, $y$
and obtained exact solutions for the warp functions. The solutions may represent
exponential (inflationary) and/or oscillating (moving warp wave) behaviours according to the chosen values of the constants $C, m, \bar{\Lambda}$ and $K$. The moving solutions for warp functions may be interpreted as {\it moving
branes} along the extra dimension through the bulk.
 
We conclude that except for the RS model where the (vanishing)
reduced cosmological constant $\Lambda$ and the higher dimensional
cosmological constant $\bar{\Lambda}$ are independent of each
other, in other higher dimensional models like DGP and STM, the
reduced cosmological constant $\Lambda$ and the higher dimensional cosmological constant $\bar{\Lambda}$ are not independent of each other, rather they mutually
affect each other. This result is of particular importance
regarding the \textit{fine-tuning} problem of the cosmological
constant. In fact, One of the challenging open questions in physical community is the cosmological constant problem (vacuum catastrophe). Considering
the cosmological constant related to the vacuum energy density,
one finds the natural value for this constant is proportional to
quartic power of the Planck mass, $M^{4}_{P}$ suggested by quantum
field theory which is $120$ orders of magnitude greater than the
observational bound \cite{Weinberg}. In this regard, concerning the cosmological
constant problem, the above mentioned result about the interplay between the cosmological constants $\Lambda$ and $\bar{\Lambda}$ deserves to be scrutinized, in more detail, elsewhere.

%%%%%%%%%%%%%%%%%%

\end{document}